\documentclass{LT23auth}
\usepackage{graphicx}

\begin{document}

\begin{frontmatter}

\title{ Imaging coherent electron wave flow in a two-dimensional electron gas}

\author[address1]{B.J. LeRoy\thanksref{thank1}},
\author[address1]{A.C. Bleszynski},
\author[address2]{M.A. Topinka},
\author[address1,address2]{R.M. Westervelt},
\author[address1]{S.E.J. Shaw},
\author[address1,address3]{E.J. Heller},
\author[address4]{K.D. Maranowski},
\author[address4]{A.C. Gossard}

\address[address1]{Department of Physics, Harvard University,
Cambridge, MA 02138 USA}

\address[address2]{Division of Engineering and Applied Sciences, Harvard 
University, Cambridge, MA 02138 USA }

\address[address3]{Department of Chemistry and Chemical Biology, Harvard 
University, Cambridge, MA 02138 USA }

\address[address4]{Materials Department, University of California, Santa Barbara, 
CA 93106 USA }

\thanks[thank1]{Corresponding author. E-mail: leroy@physics.harvard.edu}

\begin{abstract}
We measure the energy distribution of electrons passing through a two-dimensional 
electron gas using a scanning probe microscope.  We present direct spatial images 
of coherent electron wave flow from a quantum point contact formed in a 
GaAs/AlGaAs two-dimensional electron gas using a liquid He cooled SPM.  A negative 
voltage is placed on the tip, which creates a small region of depleted electrons 
that backscatters electron waves. Oscillating the voltage on the tip and locking 
into this frequency gives the spatial derivative of electron flow perpendicular to 
the direction of current flow.  We show images of electron flow using this method.  
By measuring the amount of electrons backscattered as a function of the voltage 
applied to the tip, the energy distribution of electrons is measured.
\end{abstract}

\end{frontmatter}

Scanning probe microscopy (SPM) has become increasingly important for 
understanding mescoscopic phenomena including quantum point contacts, the quantum 
Hall effect and carbon nanotubes 
\cite{Topinka2000,Topinka2001,LeRoy,Crook,Finkelstein,Woodside}. SPM techniques 
provide information on a local scale, which is not available from bulk transport 
measurements.  In this paper, we present images showing coherent electron flow 
from a quantum point contact (QPC).  We demonstrate a technique for imaging the 
spatial derivative of electron flow perpendicular to the direction of current flow 
by oscillating the voltage on the tip and locking into this frequency.  We also 
measure the distribution of electrons as a function of energy by measuring the 
number of electrons backscattered as a function of the voltage applied to the tip.

\begin{figure}[btp]
\begin{center}\leavevmode
\includegraphics[width=0.9\linewidth]{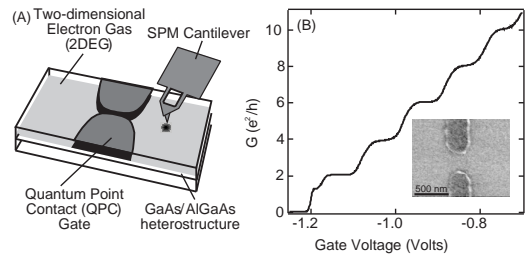}
\caption{ 
(A) Schematic diagram of the measurement setup used to image electron flow.  (B) 
Quantum point contact conductance versus gate voltage showing conductance 
plateaus.  The inset is an SEM picture of the device.
}\label{schematic}\end{center}\end{figure}

Fig. 1(A) shows the measurement setup used to image coherent electron wave flow.  
The sample used is a GaAs/AlGaAs heterostructure with a two-dimensional electron 
gas (2DEG) located 57 nm beneath the surface.  Metal gates are deposited on the 
surface to form the QPC.  A negative voltage is put on the SPM tip with respect to 
the 2DEG, which creates a small depleted region directly below the tip.  The 
depleted region can backscatter electrons reducing the conductance through the 
QPC.  By raster scanning the tip over the sample and recording the conductance, an 
image of electron flow is obtained.  Fig. 1(B) is a plot of the QPC conductance G, 
versus its width controlled by the voltage on the gate.  Well defined conductance 
plateaus are visible at multiples of 2e$^2$/h \cite{vanWees}.  The inset shows an 
SEM micrograph of the device.

\begin{figure}[btp]
\begin{center}\leavevmode
\includegraphics[width=0.8\linewidth]{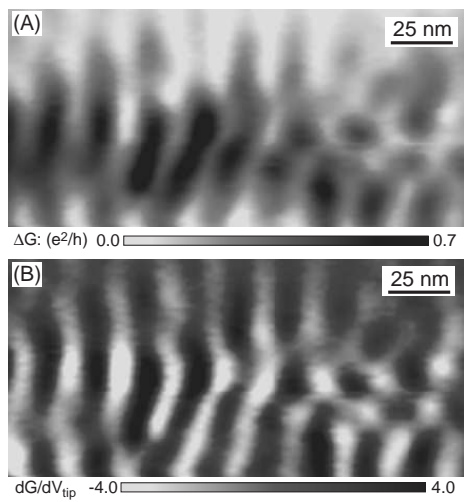}
\caption{ 
(A) Image of electron flow taken with a constant DC voltage on the tip.  Light 
areas are areas of high electron flow and dark areas have little or no electron 
flow.  (B) Image of the differential conductance with respect to tip voltage taken 
at the same location.  This images the spatial derivative of electron 
flow.}\label{comparision}\end{center}\end{figure}

Fig. 2 shows images of electron flow taken simultaneously with the two different 
measurement techniques.  Fig. 2(A) is an image of how the conductance, G changes 
as a function of tip position.  Areas of high $\Delta$G correspond to areas of 
high electron flow\cite{Topinka2000,Topinka2001}.  Fig. 2(B) is a spatial 
derivative of Fig. 2(A) acquired by adding a small oscillating voltage to the DC 
voltage applied to the tip and locking into that frequency.  Oscillating the tip 
voltage changes the size of the depleted region below the tip and hence the 
location where electrons are backscattered.  This technique therefore gives the 
spatial derivative of the electron flow because we are only sensitive to signals 
that change as a function of tip position.

\begin{figure}[btp]
\begin{center}\leavevmode
\includegraphics[width=0.75\linewidth]{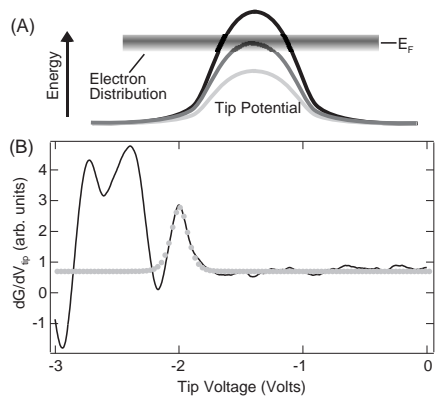}
\caption{ 
(A) Shows the effect of tip voltage on the backscattering of electrons.  The 
bottom  and top curve have little change in backscattering with tip voltage.  For 
the middle curve, the tip potential is peaked at the Fermi energy giving a large 
change in conductance with tip voltage. (B) Measurement of the differential 
conductance with respect to tip voltage as a function of tip voltage.  The curve 
near -2 Volts measures the energy distribution of the electrons.  The gray circles 
are the expected thermal distribution of electrons, which agrees very well with 
the measurement.
}\label{spectroscopy}\end{center}\end{figure}

Fig. 3 shows how we can use the tip voltage to probe the distribution of electrons 
in the 2DEG.  The tip reduces the conductance through the QPC only when the 
voltage on it is sufficient to deplete the electrons directly below and 
backscatter them\cite{Topinka2002}.  There is a distribution of electron energies 
impingent on the tip, requiring a different tip voltage to backscatter the 
different energy electrons.  Fig. 3(A) shows the tip potential for three different 
tip voltages.  As the voltage increases, the tip backscatters electrons of 
increasing energy.  For the lowest and highest tip potential, there is little 
change in the conductance with tip voltage since the distribution of electrons is 
not changing quickly.  In contrast, when the tip potential is at the Fermi energy 
there is the largest change in conductance with tip voltage since there is a 
largest change in the electron distribution at this energy.  Fig. 3(B) shows the 
differential conductance with respect to tip voltage as a function of tip voltage.  
For a thermal distribution of electrons, we would expect the signal to be the 
derivative of the Fermi function with respect to energy.  This is shown by the 
gray circles in Fig. 3(B) which fit the data very well, indicating that we are 
measuring the distribution of electrons.  The oscillations in the signal for 
larger tip voltages come from the size of the depleted region under the tip 
growing and the position of the backscattered electrons changing.  The tip is just 
moving through the interference fringes seen in Fig. 2.


%
%

\end{document}